\documentclass[12pt,letterpaper]{article}
\usepackage[utf8]{inputenc}
\usepackage{amsmath}
\usepackage{mathptmx, courier,textcomp}
\usepackage{amsfonts}
\usepackage{amssymb}
\usepackage{graphicx}
\usepackage{cite}
\usepackage{ulem}
\usepackage{caption}
\usepackage{epstopdf}
\usepackage[top=1in,bottom=1in,left=1in,right=1in]{geometry} 
\usepackage{float}
\usepackage{hyperref}
\usepackage{booktabs}
\usepackage[export]{adjustbox}
\usepackage{float}
\usepackage{multirow}
\usepackage{pifont}
\usepackage[ ]{authblk}
\usepackage{chemarrow}
\usepackage{comment}
\usepackage{makecell}
\UseRawInputEncoding
\hypersetup{
    colorlinks=true,
    linkcolor=blue,
    filecolor=magenta,      
    urlcolor=blue,
    pdftitle={Overleaf Example},
    pdfpagemode=FullScreen,
}
\begin{document}
%
%
\title{Enhanced Hydrogen Evolution Using $\beta$-MnO$_2$ Monolayer on Ni Electrode with Engineered Oxygen Vacancies}  
%
%

%
%
\author[1,$\dag$]{Faysal Rahman}
\author[1,$\dag$]{Abdul Ahad Mamun}
\author[1]{Auronno Ovid Hussain}
\author[1,*]{Muhammad Anisuzzaman Talukder}
\affil[1]{\small{Department of Electrical and Electronic Engineering \\ 

Bangladesh University of Engineering and Technology \\

Dhaka 1205, Bangladesh\\}}
\affil[$\dag$]{\small{\it{These authors contributed equally to this work.}}}
\affil[*]{\small{\it{anis@eee.buet.ac.bd}}}
%
%
\date{}
\maketitle 
\sloppy 
%
%
\begin{abstract}
Developing cost-effective and high-performance electrodes is critical for advancing hydrogen (H$_2$) production through electrochemical water splitting. In this study, we present a novel electrode design by depositing a $\beta$-MnO$_2$ monolayer on a conventional Ni(100) substrate (MnO$_2$(110)/Ni(100)) and systematically investigate its electrocatalytic properties. This work uniquely explores the influence of oxygen vacancies (OVs) at distinct sites---Osub-top and bridge sites---on both the hydrogen evolution reaction (HER) and oxygen evolution reaction (OER). Our findings reveal that the Osub-top vacancy (OOV-MnO$_2$(110)/Ni(100)) significantly enhances HER activity, achieving a hydrogen Gibbs free energy ($\Delta G_{\rm H}$) of $-0.015$ eV, which surpasses the performance of noble metals such as Pt/C ($-0.082$ eV) and Ir ($-0.08$ eV). Additionally, the cathodic exchange current density $(J_{0,c})$ of OOV-MnO$_2$(110)/Ni(100) reaches $10^{-0.774}$ Acm$^{-2}$, outperforming Pt/C ($10^{-0.92}$ Acm$^{-2}$) and Ir ($10^{-1.44}$ Acm$^{-2}$). Electrochemical analysis confirms a cathodic activation overpotential $(\eta_{a,c})$ of 0.141 V at 10 mAcm$^{-2}$ in a 0.5 M H$_2$SO$_4$ solution, achieving a hydrogen production rate (HPR) of 0.91 mmolh$^{-1}$cm$^{-2}$ at an applied voltage ($V_{\rm app}$) of 1.60 V. This study provides the first comprehensive analysis of site-specific oxygen vacancy effects on bifunctional MnO$_2$-based electrodes, demonstrating superior HER activity while maintaining dual functionality for both cathodic and anodic processes. Our results highlight the potential of engineered oxygen vacancies to develop low-cost, high-efficiency electrodes for sustainable hydrogen production, offering a competitive alternative to precious metal-based catalysts.
\end{abstract}
%
%

%
%

%
%
\section{Introduction}
Electrochemical water splitting is a promising method for producing sustainable hydrogen (H$_2$) fuel, which is crucial for reducing reliance on fossil fuels and mitigating greenhouse gas emissions \cite{ng2024elevating, reda2024green}. This process involves two half-cell reactions: the hydrogen evolution reaction (HER) at the cathode and the oxygen evolution reaction (OER) at the anode. Despite its environmental benefits, the practical application of water splitting is hindered by the sluggish kinetics of these reactions, particularly the OER, and the high energy input required to drive them efficiently. To overcome these challenges, there is an urgent need for cost-effective and high-performance electrode materials that can facilitate both HER and OER with minimal energy losses.

Noble metal-based catalysts, such as platinum (Pt) for HER and iridium (Ir) for OER, are widely regarded as the most effective materials due to their superior electrocatalytic activity and low overpotentials \cite{kazemi2024metal, tong2023precision}. However, their high cost and limited availability pose significant barriers to large-scale implementation. As a result, researchers have focused on developing alternative materials that are both abundant and capable of delivering comparable catalytic performance. Transition metal oxides (TMOs) have emerged as promising candidates due to their natural abundance, chemical stability, and tunable electronic properties \cite{vs2024harnessing}. Among these, manganese dioxide (MnO$_2$) is particularly attractive due to its low cost, environmental compatibility, and diverse structural polymorphs that offer multiple active sites for catalytic reactions \cite{du2024advance, yang2021mno2}.

MnO$_2$ exists in several crystallographic forms ($\alpha$, $\beta$, $\gamma$, and $\delta$), each with distinct structural and electronic properties \cite{wang2018improvement}. The $\beta$-phase is of particular interest due to its exceptional phase stability under both acidic and alkaline conditions, making it suitable for long-term electrochemical applications \cite{rittiruam2022first, hatakeyama2022thermal}. However, the intrinsic catalytic activity of $\beta$-MnO$_2$ is generally lower than that of other polymorphs, limiting its direct use in water-splitting applications. Recent studies suggest that the introduction of structural defects, such as oxygen vacancies (OVs), can significantly enhance the catalytic performance of TMOs \cite{li2015insight}. These vacancies can modulate the electronic structure, increase the density of active sites, and improve charge transfer kinetics, all of which are critical for enhancing HER and OER activities.

The interface between TMOs and metal substrates plays a crucial role in determining the overall electrocatalytic performance \cite{li2015insight, du2024advance}. Combining a functional material like MnO$_2$ with a conductive metal substrate can leverage the synergistic effects of both components, enhancing electron transport and catalytic activity \cite{xiong2019interface}. Nickel (Ni) is an attractive choice for such a substrate due to its low cost, good electrical conductivity, and compatibility with TMO deposition. However, the electrocatalytic performance of Ni electrodes alone is limited, especially for HER, necessitating surface modification strategies to enhance their functionality.

In this study, we present a novel electrode design by depositing a monolayer of $\beta$-MnO$_2$ on a Ni(100) substrate (MnO$_2$(110)/Ni(100)) and systematically investigate the impact of site-specific oxygen vacancies on both HER and OER performance. This research uniquely focuses on the effect of oxygen vacancies at two distinct locations---Osub-top and bridge sites---and their influence on reaction kinetics. We employ density functional theory (DFT) to analyze the electronic properties, adsorption energies, and Gibbs free energy of adsorption ($\Delta G$) to provide a comprehensive understanding of the catalytic mechanisms. Our findings reveal that the presence of Osub-top oxygen vacancies (OOV) in the MnO$_2$(110)/Ni(100) electrode significantly enhances HER activity, achieving a remarkable Gibbs free energy for H$_2$ ($\Delta G_{\rm H}$) of $-0.015$ eV, which is closer to the thermodynamic ideal than Pt/C ($-0.082$ eV) and Ir ($-0.08$ eV). This electrode also exhibits a higher cathodic exchange current density $(J_{0,c})$ of $10^{-0.774}$ Acm$^{-2}$, surpassing traditional noble metal catalysts. Furthermore, electrochemical analyses demonstrate that the OOV-MnO$_2$(110)/Ni(100) electrode achieves a cathodic activation overpotential ($\eta_{a,c}$) of 0.141 V at current density ($J$) of 10 mAcm$^{-2}$ and a hydrogen production rate (HPR) of 0.91 mmolh$^{-1}$cm$^{-2}$, outperforming pure Ni and noble metal-based counterparts.

This work makes a significant scientific contribution by providing the first systematic investigation of the site-specific effects of oxygen vacancies in an MnO$_2$ monolayer on Ni electrodes. It highlights the potential of oxygen vacancy engineering to develop low-cost, high-efficiency bifunctional electrodes capable of facilitating both HER and OER processes. The findings open new avenues for designing advanced electrode materials for sustainable hydrogen production, offering a competitive and scalable alternative to precious metal catalysts.
%
%

%
\begin{figure}[htb]
    \centering
    \includegraphics[width =0.92\linewidth]{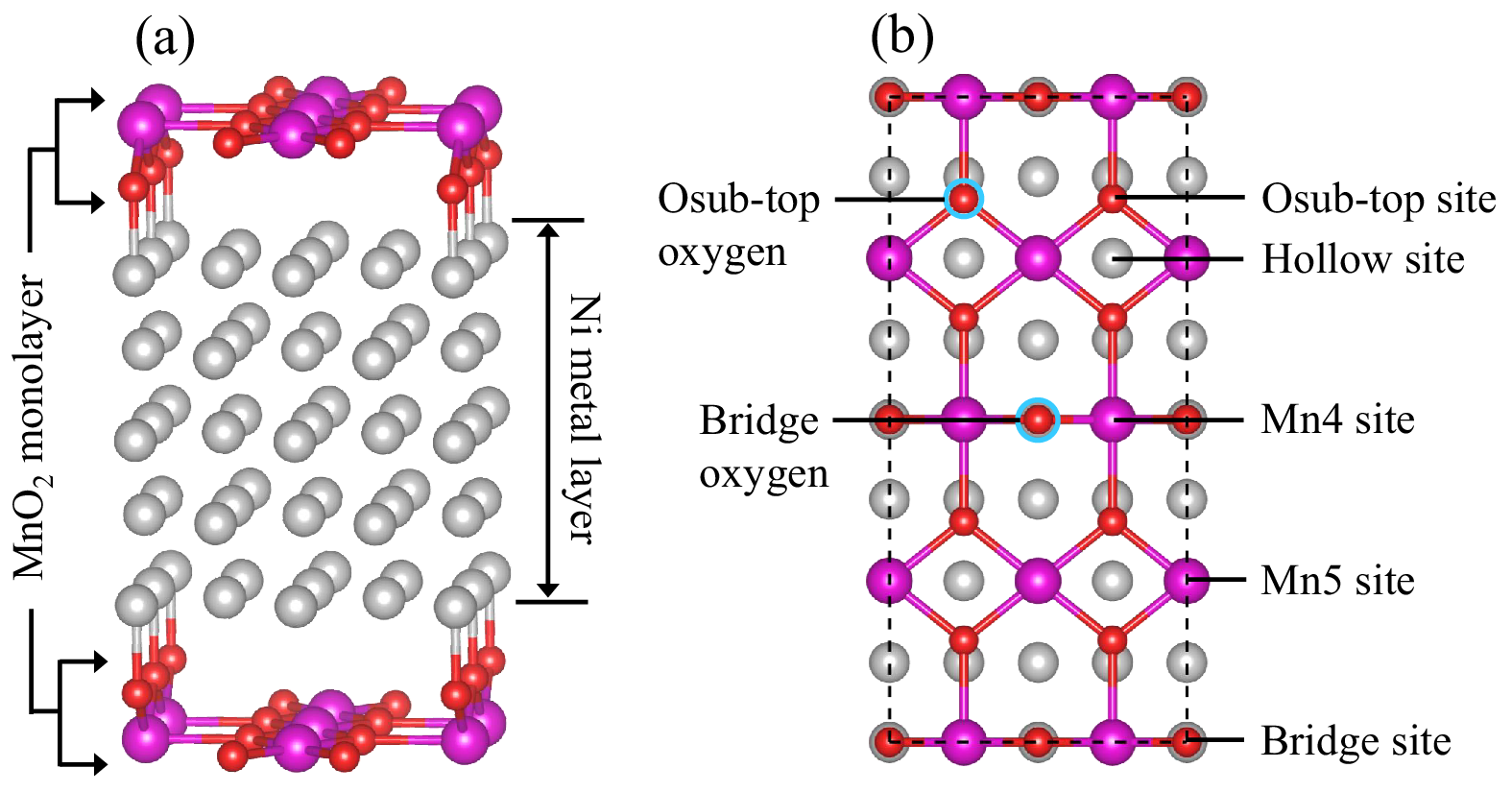}
    \caption{(a) A monolayer of $\beta$-MnO$_2$ (MnO$_2$(110)) deposited on Ni electrode structure. (b) Different adsorption and oxygen vacancy (OV) sites on the designed electrode surface. The gray, red, and purple spheres represent Ni, O, and Mn atoms respectively.}
    \label{Fig1}
\end{figure}
%
%

%
%
\section{Computational Methodology}

We performed all first-principles simulations using the self-consistent ab initio method within the framework of DFT, implemented in the Quantum Espresso (QE) package \cite{giannozzi2009quantum}. The generalized gradient approximation (GGA) with the Perdew-Burke-Ernzerhof (PBE) functional was used to model the exchange-correlation interactions, which provides a reliable balance between accuracy and computational efficiency for TMOs \cite{giannozzi2017advanced}. Core electrons were treated using the projected augmented wave (PAW) method, while the valence electrons were expanded using Kohn-Sham (KS) single-electron orbitals in a plane-wave basis. A kinetic energy cut-off of 680 eV was applied to ensure accurate total energy calculations \cite{marsman2006relaxed, proynov2013analyzing}. The charge density cut-off was set to twelve times the wavefunction cut-off energy to capture fine details of the electron distribution. To account for long-range van der Waals interactions, which are crucial for accurately modeling interfacial systems, the DFT-D3 method by Grimme et al. was incorporated \cite{grimme2010consistent}.

A 7$\times$7$\times$1 Monkhorst-Pack $k$-point mesh was employed to sample the Brillouin zone and ensure convergence of electronic states \cite{monkhorst1976special}. We used the Marzari-Vanderbilt (MV) smearing method with a smearing width of 0.02 Ry to facilitate convergence and stabilize the electronic occupations. Structural relaxations were performed using the Broyden-Fletcher-Goldfarb-Shanno (BFGS) algorithm until the atomic forces were reduced below 0.05 eV/{\AA} and the total energy change was less than 10$^{-4}$ eV. This rigorous convergence criterion ensured accurate modeling of the optimized structures and surface interactions.

We modeled the electrode using a cubic crystal structure of Ni (space group Fm$\bar{3}$m [225]) and a tetragonal crystal structure of $\beta$-MnO$_2$ (space group P4$_2$/mnm [136]) \cite{liu2024oxygen}. A monolayer of $\beta$-MnO$_2$ was oriented along the (110) plane and deposited on a five-layer Ni(100) substrate to simulate the electrode surface, as illustrated in Fig.~\ref{Fig1}(a). To prevent spurious interactions across periodic boundaries, a vacuum region of 10 {\AA} was introduced above and below the slab. The electrode surface exhibited five distinct adsorption sites, as depicted in Fig.~\ref{Fig1}(b), which were systematically investigated to evaluate their catalytic performance. For adsorption energy calculations ($E_{\rm ads}$), we considered only the adsorbates directly interacting with the electrode, as the impact of surrounding water (H$_2$O) molecules is minimal in these environments \cite{conway2000relation}.

To examine the effect of OVs on the electrocatalytic properties, we created vacancies at two specific locations: the Osub-top and bridge sites (BOV). The OV formation enthalpy ($E_{f,{\rm OV}}$) was calculated using the following relation \cite{cheng2020vacancy}

%
\begin{equation}
E_{f, {\rm OV}}=\frac{1}{n_{\rm O}}(E_{\rm OV} - E_{\rm slab}+n_{\rm O}\mu_{\rm O}),
\end{equation}
%
where $E_{\rm OV}$ is the total energy of the slab with vacancies, $E_{\rm slab}$ is the energy of the pristine slab, $\mu_{\rm O}$ represents the chemical potential of a single oxygen atom, and $n_{\rm O}$ is the number of removed oxygen atoms. This formulation provides a quantitative measure of the thermodynamic stability and likelihood of vacancy formation under experimental conditions.

We evaluated the electrocatalytic activity by calculating the hydrogen adsorption energy ($\Delta E_{{\rm ads,H}}$) for the HER and the oxygen adsorption energy ($\Delta E_{{\rm ads,O}}$) for the OER considering the electrolyte as an acidic medium. The hydrogen ion (H$^{+}$) accepts a single electron to produce adsorbed hydrogen (H$^{*}$), which is an essential step in ultimately generating H$_2$ gas in an acidic solution \cite{chabira2024comprehensive}. On the other hand, H$_2$O oxidizes to produce oxygen gas (O$_2$) through a four-step process that involves the transfer of four electrons \cite{shi2021electrochemical}. Among these four steps, the bonding interaction between the O atom and the electrode surface is crucial for understanding OER process for producing O$_2$ \cite{li2022oxygen}. A lower bonding interaction indicates that OER requires a smaller applied voltage ($V_{\rm app}$) to complete H$_2$O oxidation, while a higher bonding interaction implies that a larger $V_{\rm app}$ is necessary \cite{norskov2005trends,yang2021revisiting}. This interaction is critically dependent on $\Delta E_{{\rm ads,O}}$, so we focused on the O adsorption on the designed electrode surface to assess OER activity. The $\Delta E_{{\rm ads,H}}$ and $\Delta E_{{\rm ads,O}}$ can be written as \cite{yang2021revisiting}
%
\begin{subequations}
\begin{align}
    & \Delta E_{{\rm ads,H}} = \frac{1}{n}  (E_{\rm{slab/nH^*}} - E_{\rm{slab}} - \frac{n}{2} E_{\rm{H_2}} ),\\
    & \Delta E_{{\rm ads,O}} = \frac{1}{n}  (E_{\rm{slab/nO^*}} - E_{\rm{slab}} - \frac{n}{2} E_{\rm{O_2}} ), 
\end{align}
\end{subequations}
%
where $E_{\rm slab/nH^*}$ and $E_{\rm slab/nO^*}$ represent the total energies of the system with adsorbed H and O atoms, respectively. $E_{\rm H_2}$ and $E_{\rm O_2}$ are the reference energies of isolated hydrogen and oxygen molecules, and $n$ is the number of adsorbed species.

The parameter $\Delta G$ was calculated to account for zero-point energy corrections ($\Delta E_{\rm ZPE}$) and entropic contributions, which are essential for accurately modeling reaction thermodynamics \cite{yang2021revisiting}
%
\begin{subequations}
\begin{align}
    & \Delta G_{\rm H} = \Delta E_{\rm ads,H} + \Delta E_{\rm ZPE} - T \Delta S_{\rm H_2}, \\
    & \Delta G_{\rm O} = \Delta E_{\rm ads,O} + \Delta E_{\rm ZPE} - T \Delta S_{\rm O_2} , 
\end{align}
\end{subequations}
%
where $T$ is the temperature, and $\Delta S_{\rm H_2}$ and $\Delta S_{\rm O_2}$ are the entropy changes of H$_2$ and O$_2$ in the gas phase. We adopted the experimental values $\Delta E_{\rm ZPE} - T \Delta S_{\rm H_2} = 0.24$ eV and $\Delta E_{\rm ZPE} - T \Delta S_{\rm O_2} = 0.34$ eV \cite{norskov2005trends}.

By employing $\Delta G$, we calculated exchange current density ($J_0$) for both HER and OER using the advanced kinetics model given by \cite{mamun2024enhancing}
%
\begin{subequations} 
\begin{align}
J_0 = e k_0^\alpha \exp(-\Delta G_{\rm M}) C^{\beta}_{\rm M} P^{\gamma}_{\rm M}  \frac{\exp(-\Delta G_{\rm M}/k_{\rm B} T)}{1+\exp(-\Delta G_{\rm M}/k_{\rm B} T)}C_{\rm tot} \hspace{2cm} \text{for $\Delta G_{\rm M} >$ 0 }, \\
J_0 = e k_0^\alpha \exp(-\Delta G_{\rm M}) C^{\beta}_{\rm M} P^{\gamma}_{\rm M}  \frac{1}{1+\exp(-\Delta G_{\rm M}/k_{\rm B} T)}C_{\rm tot} \hspace{2cm} \text{for $\Delta G_{\rm M} <$ 0 }, 
\end{align}
\end{subequations}
%
where $e$ is the elementary charge of an electron, $C_{\rm tot}$ is the concentration of the total adsorption sites, $k_{\rm B}$ is the Boltzmann constant, and $k_{\rm 0}$ is the universal rate constant, whose value is 200 s$^{-1}$site$^{-1}$. The parameters $C_{\rm M}$ and $P_{\rm M}$ represent the electrolyte concentration and the system pressure, respectively. $\alpha$, $\beta$, and $\gamma$ are the empirical parameters associated with the rate constant, concentration, and system pressure, respectively. The values for these parameters were employed from Ref.~\citeonline{mamun2024enhancing}. The subscript M indicates the adsorbate atom, which is H and O. When M is considered an H (O) atom, we obtain $J_{0,c}$ ($J_{0,a}$) for HER (OER) from Eq.~4.

The electrochemical performance was further analyzed through the calculation of cathodic and anodic overpotentials ($\eta_{a,c}$ and $\eta_{a,a}$) using the activation overpotential ($\eta_a$) model described in Ref.~\citeonline{mamun2023effects}. Electrochemical impedance spectroscopy (EIS) simulations were conducted using the COMSOL Multiphysics software package \cite{mayen2019electrochemical}, with an AC voltage amplitude of 5 mV and a frequency range of 0.01 Hz to 100 kHz. We also examined polarization curves (PCs) via the Butler-Volmer equation and estimated the HPR and oxygen production rates (OPR) using Faraday's law, assuming 100\% Faradaic efficiency \cite{dickinson2020butler, mamun2023effects}.     
%
%
%
%
\section{Results and Discussion}
The designed electrode, with and without the presence of OVs, requires a comprehensive investigation to understand their electronic properties and electrocatalytic activity for both HER and OER in electrochemical water splitting. Firstly, we presented the structural characteristics and electronic properties, including $E_{f,{\rm OV}}$, density of states (DOS), work function ($\phi$), and Bader charge analysis. Next, we investigated $\Delta G$ for both H and O atoms to calculate $J_0$, which helps to study the mechanism of reaction kinetics. Finally, several electrochemical analyses, such as $\eta_a$, EIS, polarization curve (PC), and H$_2$ production, were conducted to reveal the full potential of the designed electrodes for electrochemical water-splitting application. 

%
\subsection{Structural Characteristics and Electronic Properties}
To verify the structural stability of the electrode, we optimized the crystal structures of Ni and $\beta$-MnO$_2$. The calculated lattice constant of Ni was 3.482 {\AA}, which is consistent with the experimental value of 3.52 {\AA} \cite{mahmoud2018study}, while the $\beta$-MnO$_2$ monolayer exhibited optimized lattice parameters of $a = b = 4.46$ {\AA} and c = 2.94 {\AA}, aligning well with previously reported values \cite{wang2013beta}. This strong agreement validates the accuracy of our computational approach.

The feasibility of incorporating OVs into the structure critically depends on thermodynamic phase stability, which is related to $E_{f,{\rm OV}}$. A positive value of $E_{f,{\rm OV}}$ signifies that an external fabrication process is necessary to induce the OVs, while a negative value implies that the OVs can form spontaneously within the structure. For the OOVs and BOVs in the designed electrode, we found $E_{f,{\rm OV}}$ of $1.972$ and $3.865$ eV, respectively. Both positive values strongly indicated that the designed electrode was stable without the presence of OVs. 

Various techniques can be employed to create OVs in a structure, including ion doping, ultraviolet light irradiation, thermal treatment, and chemical reduction \cite{zhuang2020oxygen}. Among these, thermal treatment and chemical reduction are the most effective and widely used methods for generating OVs in MnO$_2$ \cite{yang2024research}. Thermal treatment involves a calcination process where the concentration of OVs can be precisely controlled by adjusting the calcination temperature. At elevated temperatures, the increased atomic vibrations weaken the bonds between O and the crystal lattice, facilitating the release of O atoms and increasing the OV density.

Chemical reduction, on the other hand, utilizes a reducing agent to selectively remove surface O atoms, creating OVs. This method allows precise control over vacancy concentration by varying the duration and intensity of the reduction process, where prolonged treatment typically results in a higher OV density. Both thermal treatment and chemical reduction offer practical, scalable, and efficient approaches to engineer OVs in MnO$_2$. These methods are crucial for optimizing the electrocatalytic performance of the designed electrode by enhancing charge transfer and increasing the number of active sites.
%

%
\begin{figure}[htb]
    \centering
    \includegraphics[width =0.85\linewidth]{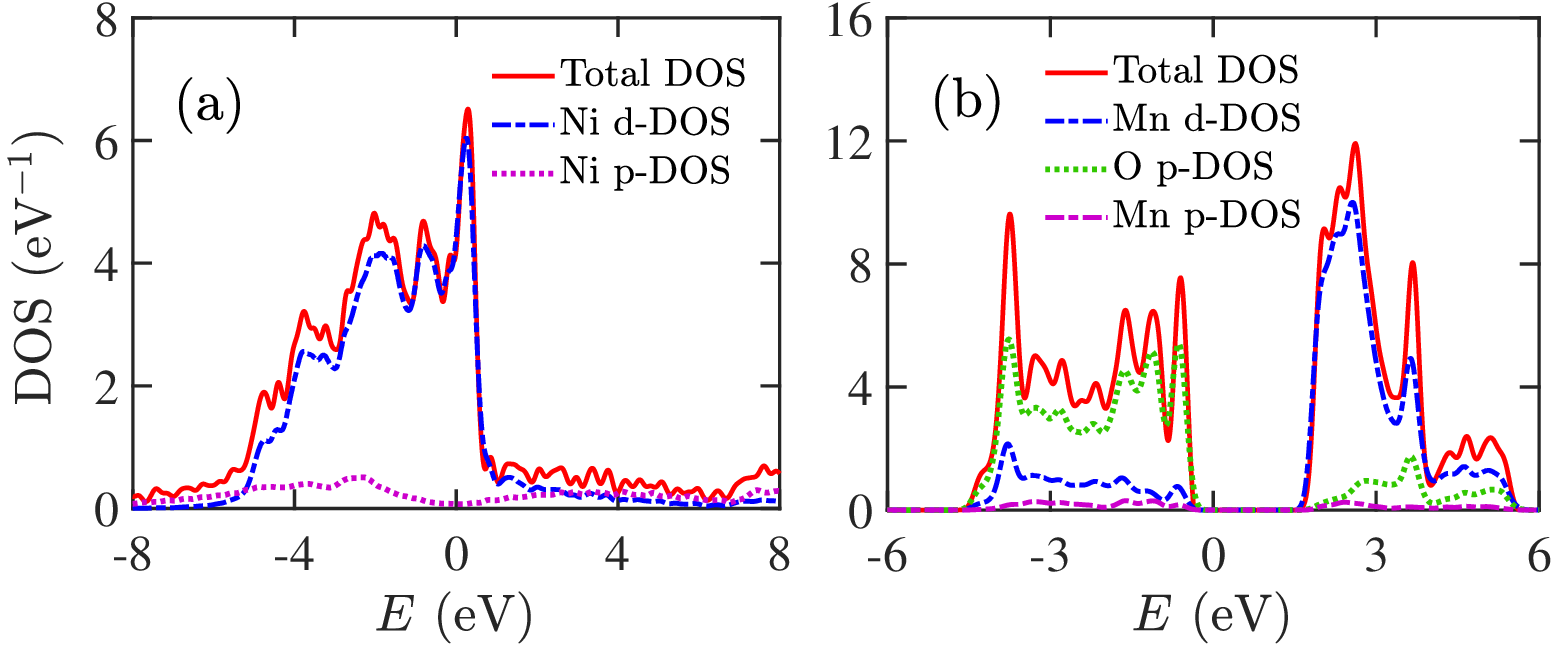}
    \caption{Density of states (DOS) for (a) Ni(100) and (b) monolayer of $\beta$-MnO$_2$ (MnO$_2$(110)) structures. For both of them, the Fermi level is set to zero.}
    \label{Fig2}
\end{figure}
%
%

The DOS provides insight into the material's electronic structure and the degeneracy of charge carriers, which helps qualitatively describe electrical conductivity. In metals, the DOS reveals no band gap energy ($E_g$), where semiconductors exhibit an $E_g$ with the conduction band (CB) and valence band (VB) positioned to the right and left of the Fermi level, respectively. A broad energy range in the DOS with heightened intensity indicates an increased number of conduction electrons, which leads to enhanced electrical conductivity and boosted cell current density ($J_{\rm cell}$) under a smaller $V_{\rm app}$. Figure \ref{Fig2}(a) presents the DOS of Ni(100), which aligns well with the findings reported by Kapcia et al., thereby affirming the reliability of our simulation approach \cite{kapcia2024ultrafast}. The d-orbital DOS of Ni(100) significantly contributes to the total DOS, indicating that the conduction electrons originate from the d-orbital. 

Figure \ref{Fig2}(b) illustrates the DOS of the MnO$_2$(110) structure, revealing a $E_g$ of approximately 2.0 eV, classifying it as a semiconductor. This value aligns closely with the experimentally measured band gap of $\beta$-MnO$_2$ ultra-thin layers, which ranges from 1.82 to 1.98 eV \cite{makhlouf2018preparation}. The CB of MnO$_2$(110) is primarily composed of Mn d-orbitals, while the VB mainly consists of O p-orbitals. The carrier degeneracy is more pronounced in the VB due to the ease of excitation of p-orbital electrons, resulting in the Fermi level shifting downward toward the VB. This electronic configuration suggests that an ultra-thin MnO$_2$(110) monolayer, when deposited on metallic Ni(100), can significantly enhance the availability of conduction electrons. This hybrid structure also provides diverse adsorption sites on the surface, making it a promising candidate for catalytic applications.

\begin{figure}[htb]
    \centering
    \includegraphics[width =1.0\linewidth]{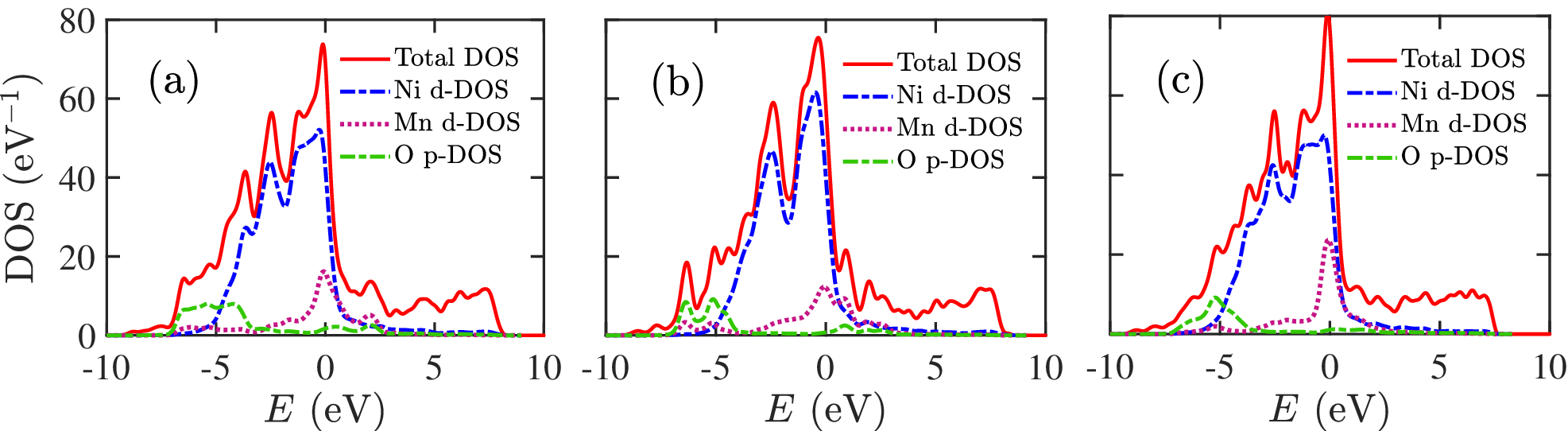}
    \caption{Density of states (DOS) for (a) MnO$_2$(110)/Ni(100), (b) OOV-MnO$_2$(110)/Ni(100), and (c) BOV-MnO$_2$(110)/Ni(100) structures. For all of them,
    the Fermi level is set to zero.}
    \label{Fig3}
\end{figure}
%
%

Figure \ref{Fig3}(a) presents the total DOS of the MnO$_2$(110)/Ni(100) structure, which shows increased intensity over a broad energy range. This enhancement reflects a higher density of conduction electrons, contributing to improved electrical conductivity. Compared to Ni(100), the total DOS on the right side of the Fermi level is more pronounced due to the contribution of the CB of MnO$_2$(110), which spans the energy range of 2.0 to 5.8 eV. These DOS features indicate that MnO$_2$(110)/Ni(100) exhibits excellent metallic properties and superior conductivity.

Figures \ref{Fig3}(b) and \ref{Fig3}(c) show the total DOS of MnO$_2$(110)/Ni(100) after the introduction of OOVs and BOVs, respectively. In both cases, the changes in the total DOS were minimal, indicating that the presence of OVs has a negligible effect on the overall electronic structure. Slight variations were observed in the O's p-orbital and Mn's d-orbital DOS near the Fermi level, which can be attributed to the alternating oxidation states of Mn induced by the formation of OVs. Overall, the DOS analysis of MnO$_2$(110)/Ni(100), with and without OVs, highlights the structure's remarkable electronic properties. The high density of degenerate conduction electrons enhances electrical conductivity, thereby increasing $J_{\rm cell}$ while reducing the required $V_{\rm app}$ in the electrochemical water-splitting process.

The $\phi$ of the metal-catalyst electrode plays a critical role in facilitating charge carrier transfer between the electrode and the electrolyte during both the HER and the OER processes. In acidic conditions, a higher $\phi$ leads to an increased accumulation of H$^{+}$ ions in the electric double layer (EDL), accelerating the adsorption and desorption of hydrogen (H) atoms and thereby enhancing HER activity in electrochemical water splitting. Conversely, a lower $\phi$ promotes stronger adsorption of H atoms on d-block transition metals, which can hinder the desorption process and result in poor HER performance. For OER, however, a lower $\phi$ reduces the charge-transfer barrier, facilitating electron exchange across the electrode-electrolyte interface and improving overall OER efficiency.

%
%
\begin{table}[H]
\centering
\caption{ The work function ($\phi$) and Bader charge analysis 
for Ni(100) and the designed electrodes.}
\resizebox{0.98\textwidth}{!}{%
\begin{tabular}{ ccccccc }
\Xhline{3\arrayrulewidth}
     Electrode & Work Function & \multicolumn{5}{c}{Bader Charge ($e$C)} \\
     \cline{3-7}
       & $\phi$ (eV) & Ni & Mn4 & Mn5 & Osub-top O & Bridge O \\
    \Xhline{2\arrayrulewidth}
    Ni(100) & 5.124 & 0 & - & - & - & - \\
    MnO$_2$(110) & 5.022 & - & $+1.502$ & $+1.380$ & $-1.040$ & $-0.810$ \\
    MnO$_2$(110)/Ni(100) & 4.806 & $+0.025$ & $+1.331$ & $+1.262$ & $-1.006$ & $-0.851$ \\
    OOV-MnO$_2$(110)/Ni(100) & 5.403 & $-0.030$ & $+0.995$ & $+1.054$ & $-0.876$ & $-0.872$ \\
    BOV-MnO$_2$(110)/Ni(100) & 4.138 & +0.012  & +0.863 & +1.121 & $-1.055$ & - \\
    \Xhline{3\arrayrulewidth}
\end{tabular}}
\label{Table1}
\end{table}
%
%

Table \ref{Table1} presents $\phi$ and Bader charge analysis for Ni(100) and the designed MnO$_2$(110)/Ni(100) electrodes. The MnO$_2$(110)/Ni(100) electrode exhibited a lower $\phi$ compared to Ni(100), suggesting its potential for enhanced OER activity due to the reduced charge-transfer barrier. Among the structures with OVs, the OOV-MnO$_2$(110)/Ni(100) displayed a higher $\phi$, indicating that it may provide superior HER activity by promoting faster H adsorption-desorption dynamics. In contrast, the BOV-MnO$_2$(110)/Ni(100) showed a lower $\phi$ than Ni(100), suggesting that it may exhibit improved OER performance. While $\phi$ serves as a useful indicator of catalytic behavior, it is challenging to evaluate OER activity solely based on $\phi$ for electrodes with OVs. This is because the electrocatalytic performance is also strongly influenced by the concentration and spatial distribution of OVs, which modulate the electronic structure and charge-transfer characteristics.

The electronic charge property can be quantified using Bader charge analysis, which measures the ionic charge of each atom and provides a qualitative understanding of charge transfer during the water-splitting process. For MnO$_2$(110)/Ni(100), the ionic charge of the Mn atom decreased compared to that of the Mn atom in MnO$_2$(110) due to interactions between the Ni slab and MnO$_2$(110). In contrast, the changes in the ionic charge of the O atoms were negligible. After the incorporation of OVs, the ionic charge of the Mn atom significantly decreased, suggesting that oxidation of Mn may occur during the HER process, thereby enhancing electrocatalytic activity. In the case of OOV-MnO$_2$(110)/Ni(100), the ionic charge of the O atom decreased at the Osub-top site, which may affect its bonding with the adsorbate atom. For BOV-MnO$_2$(110)/Ni(100), however, the ionic charge of the O atom remained unchanged at the Osub-top site. Additionally, the absence of an O atom at the bridge site may have influenced the formation of a bond with the adsorbate O atom, making the adsorption and desorption processes significantly more challenging.           

%
\begin{table}[H]
\centering
\caption{Hydrogen adsorption energy ($\Delta E_{\rm ads,H}$) and Gibbs free adsorption energy ($\Delta G_{\rm H}$) for the designed electrodes, including adsorbed hydrogen (H) atom distance from the adsorption site.}
\resizebox{0.84\textwidth}{!}{%
\begin{tabular}{ ccccc }
\Xhline{3\arrayrulewidth}
     Electrodes & Adsorption & $\Delta E_{\rm ads,H}$ & $\Delta G_{\rm H}$ & H atom distance\\ 
      & Sites & (eV) & (eV) & (\AA) \\
    \Xhline{2\arrayrulewidth}
    \multirow{5}{10em}{\rm {MnO$_2$(110)/Ni(100)}} & Mn4 & 
        $1.422$ & $1.662$ & $1.512$ \\
      & Mn5 & $0.572$ & $0.812$ & $1.552$ \\
      & Hollow & $1.287$ & $1.527$ & $1.427$ \\
      & Osub-top & $-0.626$ & $-0.386$ & $0.985$ \\
      & Bridge & $0.415$ & $0.655$ & $0.980$ \\
    \Xhline{2\arrayrulewidth}
    \multirow{5}{12em}{\rm {OOV-MnO$_2$(110)/Ni(100)}} & Mn4 & $-0.021$ & $0.219$ & $1.730$ \\
      & Mn5 & $-0.255$ & $-0.015$ & $1.871$ \\
      & Osub-top & $0.054$ & $0.294$ & $0.988$ \\
      & Bridge & $0.163$ & $0.403$ & $0.982$ \\
    \Xhline{2\arrayrulewidth}
    \multirow{5}{12em}{\rm {BOV-MnO$_2$(110)/Ni(100)}} & Mn4 & $0.121$ & $0.361$ & $1.554$ \\
      & Mn5 & $0.403$ & $0.643$ & $1.568$ \\
      & Hollow & $1.354$ & $1.594$ & $1.457$ \\
      & Osub-top & $0.233$ & $0.473$ & $2.427$ \\
    \Xhline{3\arrayrulewidth}
\end{tabular}}
\label{Table2}
\end{table}
%
%

%

%
\subsection{Adsorption Energy and Evolution Reaction Kinetic Rate}

The parameters $\Delta E_{\rm ads,H}$ and $\Delta G_{\rm H}$ are essential for understanding the kinetic rate of HER activity, particularly in relation to the processes of adsorption and desorption, bonding interactions with the slab, and the reaction coordinate pathways. An ideal $\Delta G_{\rm H}$ value that is close to zero signifies excellent electrocatalytic activity in electrochemical water splitting \cite{norskov2005trends}. A positive value of $\Delta G_{\rm H}$ indicates endothermic processes, whereas a negative value signifies exothermic reactions \cite{mamun2024enhancing}. Furthermore, a positive $\Delta G_{\rm H}$ indicates a thermodynamically unfavorable adsorption and desorption process, likely due to repulsive forces between the adsorbate atom and the electrode surface.

Table \ref{Table2} presents $\Delta E_{\rm ads,H}$ and $\Delta G_{\rm H}$ for each adsorption site of the designed electrodes, along with the distance of the adsorbed H atom from the site. For MnO$_2$(110)/Ni(100), both $\Delta E_{\rm ads,H}$ and $\Delta G_{\rm H}$ were positive at all adsorption sites, except for the Osub-top site. The values for the Osub-top site were $\Delta E_{\rm ads,H} = -0.626$ eV and $\Delta G_{\rm H} = -0.386$ eV, while $\Delta G_{\rm H}$ for Ni(100) was $-0.28$ eV. This comparison suggests that MnO$_2$(110)/Ni(100) may exhibit a slower evolution rate compared to Ni(100). Remarkably, OOV-MnO$_2$(110)/Ni(100) showed smaller absolute values for both $\Delta E_{\rm ads,H}$ and $\Delta G_{\rm H}$, indicating its potential to enhance electrocatalytic activity for the HER.

For the Mn5 adsorption site, both $\Delta E_{\rm ads,H}$ and $\Delta G_{\rm H}$ were negative, while the Mn4 site displayed a small negative value for $\Delta E_{\rm ads,H}$. These values suggest a weak adsorption force, which can facilitate a more efficient desorption process for hydrogen production. However, the positive $\Delta G_{\rm H}$ for the Mn4 site limits the spontaneity of the adsorption and desorption processes. Conversely, all adsorption sites in BOV-MnO$_2$(110)/Ni(100) demonstrated positive values for both $\Delta E_{\rm ads,H}$ and $\Delta G_{\rm H}$, making this configuration unsuitable for HER in electrochemical water splitting. The distance of the adsorbed H atom ranged from 1.45 to 1.90 {\AA}, with the Osub-top and bridge sites showing lower distances of $<1$ {\AA}, excluding the BOV-MnO$_2$(110)/Ni(100) electrode. Notably, the adsorbed hydrogen deviated from the Osub-top site of BOV-MnO$_2$(110)/Ni(100) due to a non-uniform charge distribution. This deviation may result in the adsorbed H atom occupying the bridge vacancy site, leading to a significantly larger distance compared to the other sites.

%
\begin{figure}[htb]
    \centering
    \includegraphics[width =0.98\linewidth]{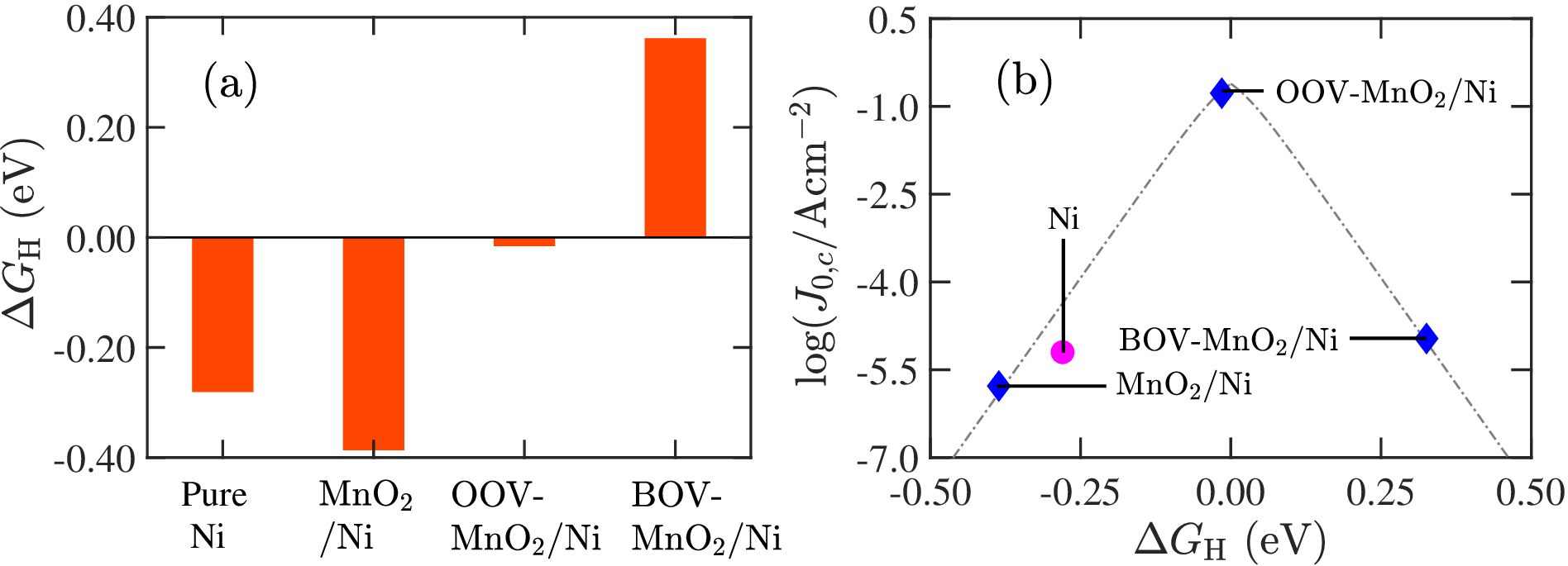}
    \caption{(a) Hydrogen Gibbs free adsorption energy ($\Delta G_{\rm H}$) of pure Ni(100) and the designed electrodes for the active sites in a preferable reaction pathway. (b) Comparison of log($J_{0,c}$/Acm$^{-2}$) vs. $\Delta G_{\rm H}$ for pure Ni(100) and the designed electrodes.}
    \label{Fig4}
\end{figure}
%
%

The active adsorption site of the designed electrodes critically depends on the total system energy (see Table S1), while the minimum energy value indicates the most favorable reaction pathway. The active sites for MnO$_2$(110)/Ni(100), OOV-MnO$_2$(110)/Ni(100), and BOV-MnO$_2$(110)/Ni(100) electrodes were identified as Osub-top, Mn5, and Mn4, respectively. Figure \ref{Fig4}(a) presents the $\Delta G_{\rm H}$ for these electrodes at their active sites. BOV-MnO$_2$(110)/Ni(100) showed a positive $\Delta G_{\rm H}$ of 0.361 eV, highlighting the challenges of using it in the HER process due to thermodynamic limitations. Remarkably, OOV-MnO$_2$(110)/Ni(100) displayed $\Delta G_{\rm H}$ of $-0.015$ eV, which is closer to zero than noble metals like Pt/C ($-0.082$ eV) and Ir ($-0.08$ eV). This value of $\Delta G_{\rm H}$ suggests that OOV-MnO$_2$(110)/Ni(100) exhibits comparable better electrocatalytic activity to these noble metals. On the other hand, MnO$_2$(110)/Ni(100) had a higher absolute value of $\Delta G_{\rm H}$ than Ni(100), leading to its inferior catalytic activity in HER. 

In addition, we calculated the $J_{0,c}$ to compare the kinetic rates for these electrodes in HER activity, as illustrated in Fig.~\ref{Fig4}(b). MnO$_2$(110)/Ni(100) displayed the lowest $J_{0,c}$, indicating slower reaction kinetic rate during the HER process. In contrast, OOV-MnO$_2$(110)/Ni(100) showed the maximum $J_{0,c}$ of 10$^{-0.774}$ Acm$^{-2}$, showcasing excellent reaction kinetic suitable for efficient H$_2$ production. As a low-cost and earth-abundant material, OOV-MnO$_2$(110)/Ni(100) outperformed noble metals and can serve as an effective and efficient cathode in a water-splitting system that may replace noble metals. 

The adsorption of O atoms is a crucial intermediate step in the oxidation of H$_2$O and plays an important role in the OER. To explore the bifunctional nature of the designed electrodes, we focused on the adsorption and desorption processes of O atoms on the electrode surfaces. Table \ref{Table3} presents the values for $\Delta E_{\rm ads,O}$, $\Delta G_{\rm O}$, and the distance of the adsorbed O atom from the electrode surface. The values of $\Delta E_{\rm ads,O}$ and $\Delta G_{\rm O}$ were negative for all electrodes, with the exception of MnO$_2$(110)/Ni(100) at the bridge site, where $\Delta E_{\rm ads,O}$ was measured at 2.09 eV, indicating a strong repulsive force during the O adsorption process. A more negative value of $\Delta E_{\rm ads,O}$ signifies a stronger tendency for a bond to form between the adsorbed O atom and the atoms of the electrode surface. 

%
%
\begin{table}[htb]
\centering
\caption{Oxygen adsorption energy ($\Delta E_{\rm ads,O}$) and Gibbs free adsorption energy ($\Delta G_{\rm O}$) for the designed electrodes, including adsorbed oxygen (O) atom distance from the adsorption site.}
\resizebox{0.84\textwidth}{!}{%
\begin{tabular}{ ccccc }
\Xhline{3\arrayrulewidth}
     Electrodes & Adsorption & $\Delta E_{\rm ads,O}$ & $\Delta G_{\rm O}$ & O atom distance\\ 
      & Sites & (eV) & (eV) & (\AA) \\
    \Xhline{2\arrayrulewidth}
    \multirow{5}{10em}{\rm {MnO$_2$(110)/Ni(100)}} & Mn4 & $-1.043$ & $-0.703$ & $1.575$ \\
      & Mn5 & $-1.583$ & $-1.243$ & $1.552$ \\
      & Hollow & $-0.714$ & $-0.373$ & $1.566$ \\
      & Osub-top & $-1.452$ & $-1.112$ & $1.692$ \\
      & Bridge & $2.090$ & $2.430$ & $1.321$ \\
    \Xhline{2\arrayrulewidth}
    \multirow{5}{12em}{\rm {OOV-MnO$_2$(110)/Ni(100)}} & Mn4 & $-1.391$ & $-1.051$ & $1.577$ \\
      & Mn5 & $-1.323$ & $-0.983$ & $1.568$ \\
      & Osub-top & $-2.291$ & $-1.951$  & $2.298$ \\
      & Bridge & $-0.739$ & $-0.399$ & $1.430$ \\
    \Xhline{2\arrayrulewidth}
    \multirow{5}{12em}{\rm {BOV-MnO$_2$(110)/Ni(100)}} & Mn4 & $-2.042$ & $-1.702$ & $1.569$ \\
      & Mn5 & $-1.985$ & $-1.645$ & $1.573$ \\
      & Hollow &$-0.717$ & $-0.377$  & $1.983$ \\
      & Osub-top & $-3.092$ & $-2.752$ & 2.458 \\
    \Xhline{3\arrayrulewidth}
\end{tabular}}
\label{Table3}
\end{table}
%
%

For OOV-MnO$_2$(110)/Ni(100), the lowest $E_{\rm ads,O}$ was found to be $-2.291$ eV, and the distance of the adsorbed O atom was significantly larger than the distances observed at other sites. This phenomenon implies a strong bond interaction and highlights the displacement of the adsorbed atom toward the OV site, facilitating structural reconstruction. Similarly, BOV-MnO$_2$(110)/Ni(100) at the Osub-top site exhibited the same behavior. This trend arises from the high concentration of OVs \cite{li2015insight}. Therefore, it is essential to carefully control the concentration of OVs to optimize the electrocatalytic activity during the OER process. In contrast, MnO$_2$(110)/Ni(100) without OVs demonstrated comparatively weak adsorption of the O atom relative to other designed electrodes, suggesting that it could be utilized as an anode to enhance OER activity in electrochemical water splitting.

%
\begin{figure}[htb]
    \centering
    \includegraphics[width =0.98\linewidth]{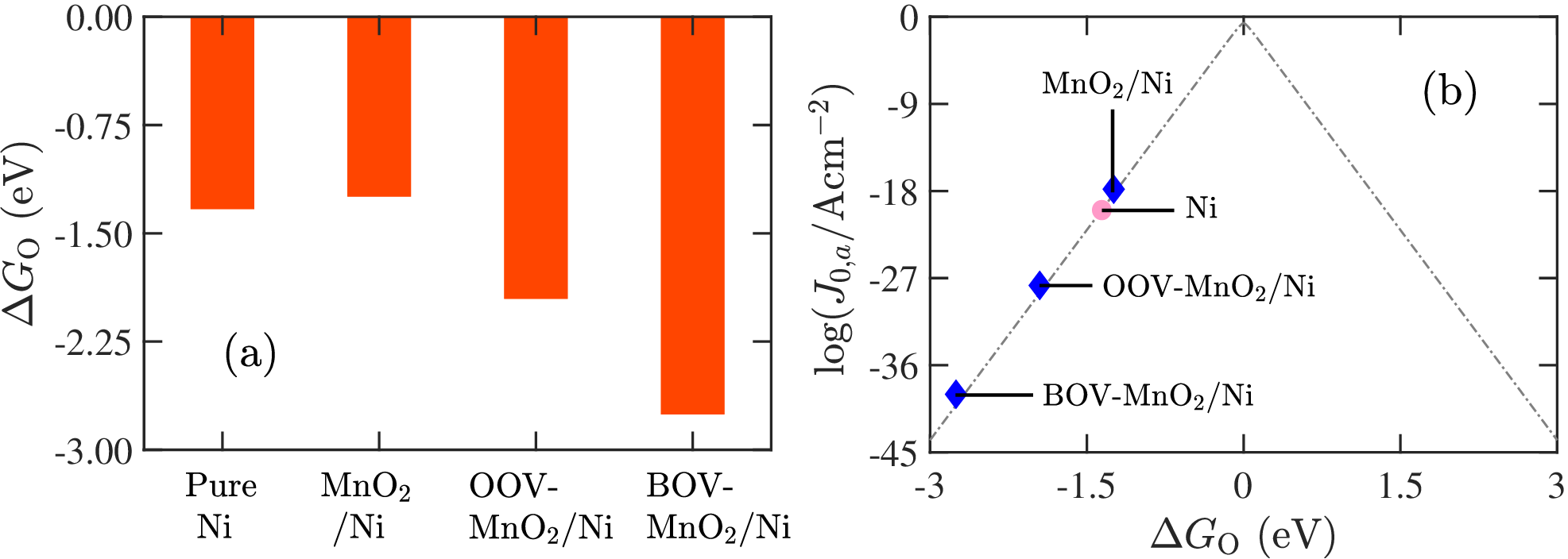}
    \caption{(a) Oxygen Gibbs free adsorption energy ($\Delta G_{\rm O}$) of pure Ni(100) and the designed electrodes for the active sites in a preferable reaction pathway. (b) Comparison of log($J_{0,a}$/Acm$^{-2}$) vs. $\Delta G_{\rm O}$ for pure Ni(100) and the designed electrodes. }
    \label{Fig5}
\end{figure}
%
%

The active site for the OER on MnO$_2$(110)/Ni(100) was identified at the Mn4 site, while the Osub-top site was the preferred active site for both OOV-MnO$_2$(110)/Ni(100) and BOV-MnO$_2$(110)/Ni(100), based on the analysis of the total system energy (see Table S2). Figure \ref{Fig5}(a) presents the values of $\Delta G_{\rm O}$ for these active sites. The $\Delta G_{\rm O}$ of MnO$_2$(110)/Ni(100) was measured at $-1.243$ eV, whereas Ni(100) had a $\Delta G_{\rm O}$ of $-1.33$ eV, which is lower than that of MnO$_2$(110)/Ni(100). This comparison indicates that the designed electrode shows better electrocatalytic activity than Ni for the OER.

However, OOV-MnO$_2$(110)/Ni(100) and BOV-MnO$_2$(110)/Ni(100) exhibited higher absolute values of $\Delta G_{\rm O}$, suggesting a tendency for stronger bonding with the electrode surface. This stronger bonding may limit effective adsorption and desorption processes. Figure \ref{Fig5}(b) illustrates the values of anodic exchange current density ($J_{0,a}$) for these electrodes, focusing on the dynamics of oxygen adsorption to provide both quantitative and qualitative insights. The $J_{0,a}$ of MnO$_2$(110)/Ni(100) showed a slight increase compared to Ni(100), indicating better reaction kinetics during the OER process. However, the lower $J_{0,a}$ values for OOV-MnO$_2$(110)/Ni(100) and BOV-MnO$_2$(110)/Ni(100) could restrict the OER process, making it nearly impossible to fully oxidize H$_2$O for BOV-MnO$_2$(110)/Ni(100).

It is noteworthy that our results reveal that MnO$_2$(110)/Ni(100) exhibited bifunctional behavior, with $J_{0,c}$ of $10^{-5.778}$ Acm$^{-2}$ for the HER and $J_{0,a}$ of $10^{-17.812}$ Acm$^{-2}$ for the OER process. Although the reaction kinetics are comparatively slower, this type of bifunctional electrode is highly effective when applying a pulsating direct current voltage in an electrochemical system. This approach can enhance cell efficiency and help prevent corrosion by disrupting the electric double layer (EDL) and the region of space charge polarization. 
%
%

%
\begin{figure}[htb]
    \centering
    \includegraphics[width =0.98\linewidth]{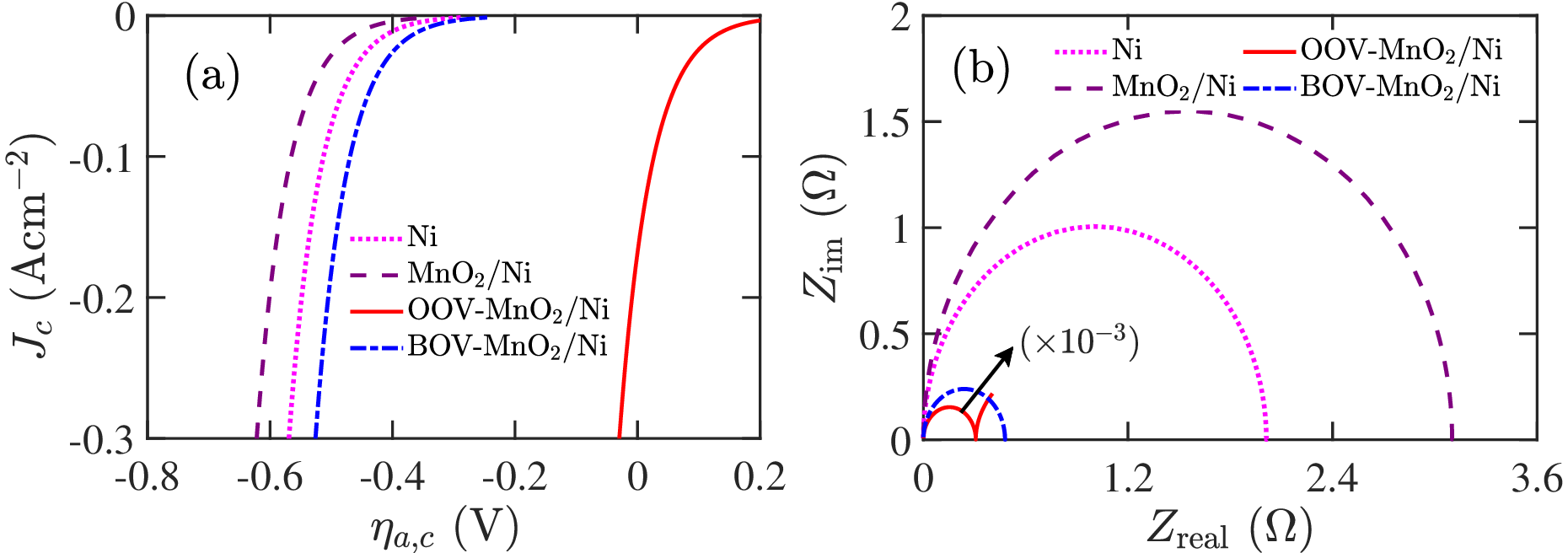}
     \caption{Analysis of (a) cathodic current density ($J_c$) vs. cathodic activation overpotential ($\eta_{a,c}$) and (b) electrochemical impedance spectroscopy (EIS) in cathodic half-cell reaction for pure Ni(100) and the designed electrodes.}
    \label{Fig6}
\end{figure}
%
%

%
\subsection{Electrochemical Performances}

For electrochemical analyses in an electrochemical system, it is crucial to evaluate the significant loss of $\eta_a$ and the charge transfer mechanism at the interface between the electrode and electrolyte, which is characterized by the charge transfer resistance ($R_{\rm CT}$). We calculated $R_{\rm CT}$ using the Nyquist plot obtained from EIS, where the diameter of the Nyquist semicircle represents the value of $R_{\rm CT}$. Figures \ref{Fig6}(a) and \ref{Fig6}(b) illustrate the $\eta_{a,c}$ and the EIS plot of the designed electrodes for the HER activity, respectively. For the Ni(100) electrode, the value of $\eta_{a,c}$ was determined to be 0.398 V at $J$ of 10 mAcm$^{-2}$ in a 0.5 M H$_2$SO$_4$ solution. The experimental value under the same conditions was 0.395 V, which closely aligns with our simulated result \cite{bakovic2021electrochemically}. 

In the case of the MnO$_2$(110)/Ni(100) electrode, the $\eta_{a,c}$ was measured at 0.451 V under the same conditions, which is higher than that of Ni(100), resulting in poorer electrochemical performance. Conversely, the $\eta_{a,c}$ for BOV-OV-MnO$_2$(110)/Ni(100) decreased slightly to 0.355 V under the same conditions. Notably, the OOV-MnO$_2$(110)/Ni(100) configuration exhibited the minimum loss of $\eta_{a,c}$ at 0.141 V due to the highest reaction kinetics, facilitating efficient hydrogen production with a lower $V_{\rm app}$.

The calculated $R_{\rm CT}$ for the Ni(100) electrode was 2.012 $\Omega$ cm$^{-2}$, considering the half-cell reaction of H$^+$ reduction. The lowest $R_{\rm CT}$ was observed for OOV-MnO$_2$(110)/Ni(100), measuring 0.0031 $\Omega$ cm$^{-2}$, which allows for rapid charge transport across the electrode-electrolyte interface. For the MnO$_2$(110)/Ni(100) electrode, $R_{\rm CT}$ was found to be higher than that of Ni(100), indicating poorer charge transport efficiency. However, $R_{\rm CT}$ for BOV-MnO$_2$(110)/Ni(100) was lower than that of Ni(100), suggesting improved charge transport for hydrogen production in electrochemical water splitting.       

%
\begin{figure}[htb]
    \centering
    \includegraphics[width =0.98\linewidth]{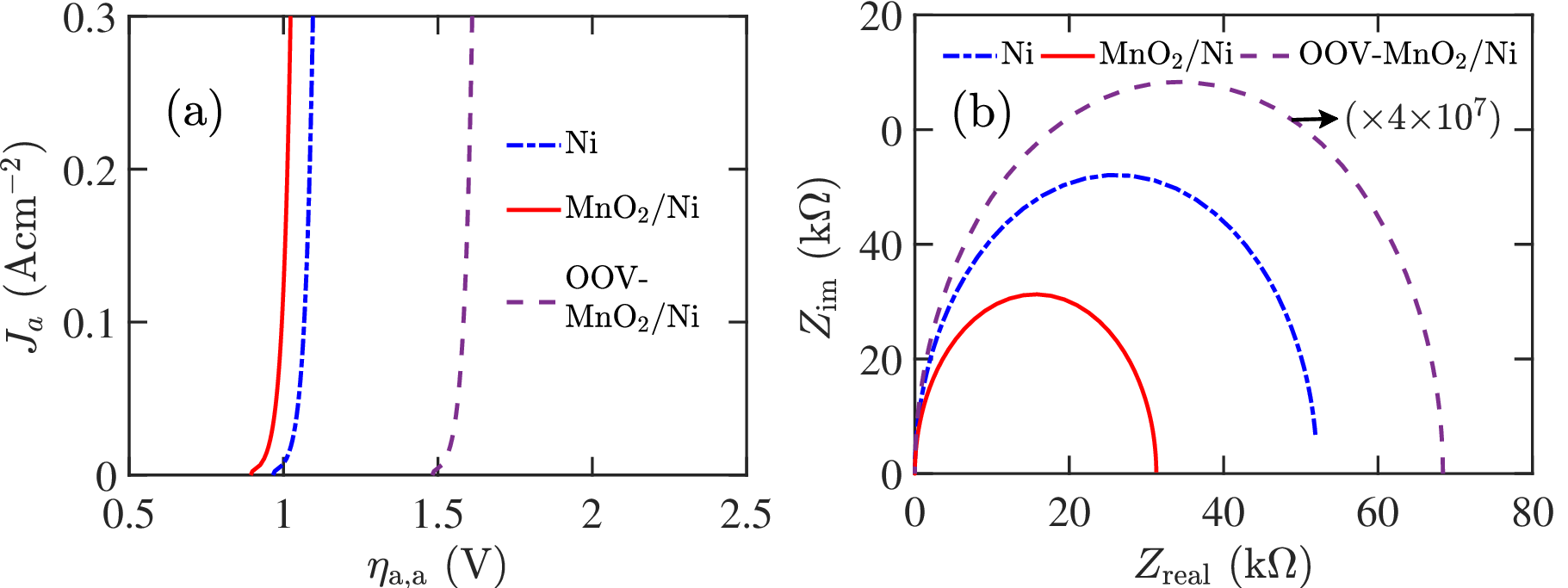}
     \caption{Analysis of (a) anodic current density ($J_a$) vs. anodic activation overpotential ($\eta_{a,a}$) and (b) electrochemical impedance spectroscopy (EIS) in anodic half-cell reaction for pure Ni(100) and the designed electrodes.}
     \label{Fig7}
\end{figure}
%
%

We also investigated the $\eta_{a,a}$ and conducted an EIS analysis for the designed electrodes, with the exception of BOV-MnO$_2$(110)/Ni(100), during the OER process, as shown in Figs.~\ref{Fig7}(a) and (b). The Ni(100) electrode exhibited a $\eta_{a,a}$ of 1.01 V at $J$ of 10 mAcm$^{-2}$ in a 0.5 M H$_2$SO$_4$ solution. In contrast, the $\eta_{a,a}$ for MnO$_2$(110)/Ni(100) was reduced to 0.937 V under the same conditions, indicating superior electrochemical performance compared to Ni(100). The OOV-MnO$_2$(110)/Ni(100) showed a significantly higher overpotential due to its lower reaction kinetics during the OER process. The $R_{\rm CT}$ for Ni(100) was 51.88 k$\Omega$, which decreased to 31.20 k$\Omega$ for MnO$_2$(110)/Ni(100), resulting in enhanced charge transport at the electrode-electrolyte interface. In contrast, the OOV-MnO$_2$(110)/Ni(100) exhibited an extremely high $R_{\rm CT}$ in the G$\Omega$ range, indicating that charge transport is limited, which hinders the completion of the OER process in electrochemical water splitting.

%
\begin{figure}[htb]
    \centering
    \includegraphics[width =0.97\linewidth]{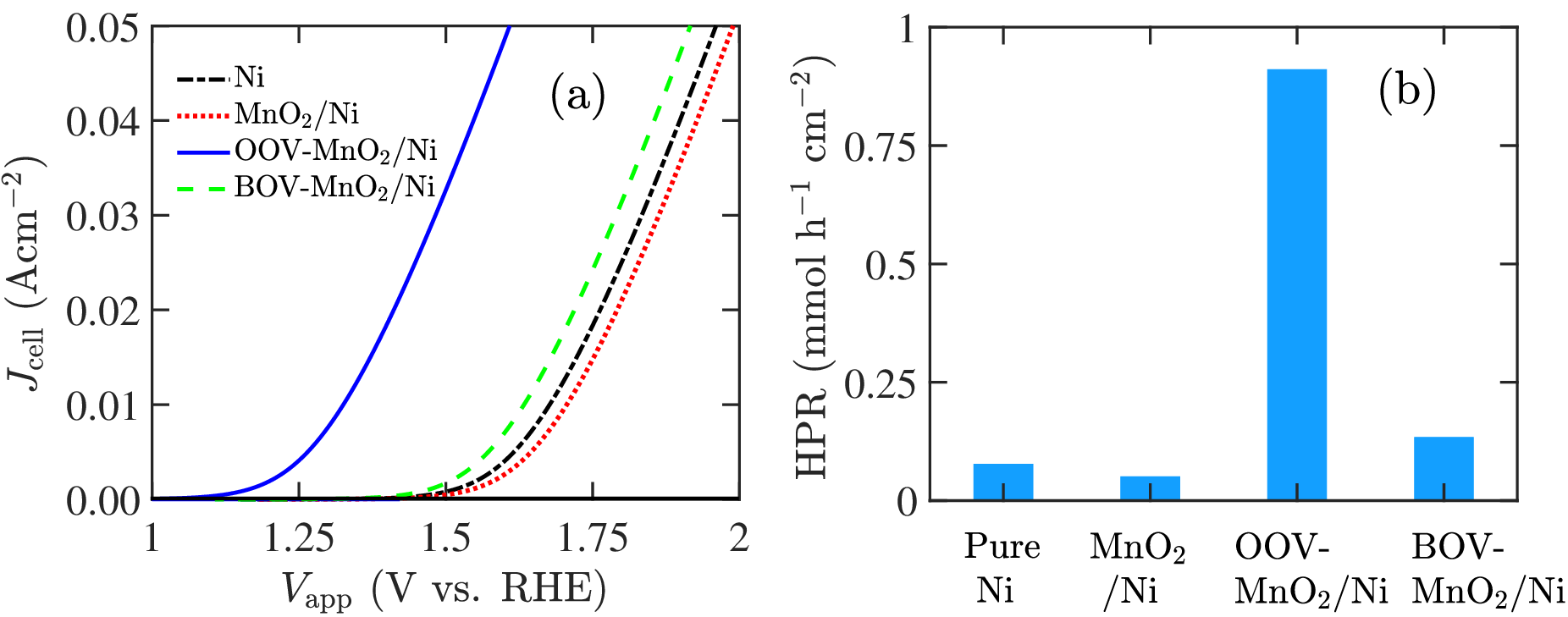}
    \caption{(a) Polarization curve (PC) and (b) hydrogen production rate (HPR) for pure Ni(100) and the designed electrodes in hydrogen evolution reaction (HER).}
    \label{Fig8}
\end{figure}
%
%

The PC analysis determines the required $V_{\rm app}$ at a specified $J$ and onset potential ($V_{\rm on}$), which is defined as the potential needed to achieve $J = 1.0$ mAcm$^{-2}$. A lower $V_{\rm on}$ for an electrode indicates that a smaller $V_{\rm app}$ is needed to initiate the evolution reactions in electrochemical water splitting. Figure \ref{Fig8}(a) illustrates the PC characteristics of the designed electrodes, taking into account both $\eta_{a,c}$ and the resistive losses from the electrolyte in a 0.5 M H$_2$SO$_4$ solution, using the reversible hydrogen electrode (RHE) as the reference electrode. The MnO$_2$(110)/Ni(100) electrode displayed less favorable PC characteristics compared to Ni(100). In contrast, both the BOV-MnO$_2$(110)/Ni(100) and OOV-MnO$_2$(110)/Ni(100) electrodes exhibited superior PC behavior, attributed to their faster reaction kinetics and enhanced charge transport efficiency.

For the electrodes Ni(100), MnO$_2$(110)/Ni(100), and BOV-MnO$_2$(110)/Ni(100), the measured $V_{\rm on}$ values were 1.513 V, 1.543 V, and 1.470 V, respectively. Notably, the OOV-MnO$_2$(110)/Ni(100) had a significantly lower $V_{\rm on}$ of 1.162 V, which is much smaller than that of Ni(100). This reduction is due to significantly decreased $\eta_{a,c}$ and $R_{\rm CT}$, resulting in more efficient hydrogen production with lower $V_{\rm app}$ in the water-splitting process.

The electrochemical performance of the HPR for the designed electrodes was evaluated at $V_{\rm app} = 1.60$ V, as shown in Fig.~\ref{Fig8}(b). The OOV-MnO$_2$(110)/Ni(100) electrode achieved the highest HPR of 0.91 mmolh$^{-1}$cm$^{-2}$, significantly surpassing that of Ni(100). Due to higher $\eta_{a,c}$ and $R_{\rm CT}$, the MnO$_2$(110)/Ni(100) electrode exhibited the lowest HPR among the tested electrodes. However, the HPR for BOV-MnO$_2$(110)/Ni(100) increased to 0.133 mmolh$^{-1}$cm$^{-2}$, which is approximately twice that of Ni(100).

%
%
\begin{figure}[htb]
    \centering
    \includegraphics[width =0.97\linewidth]{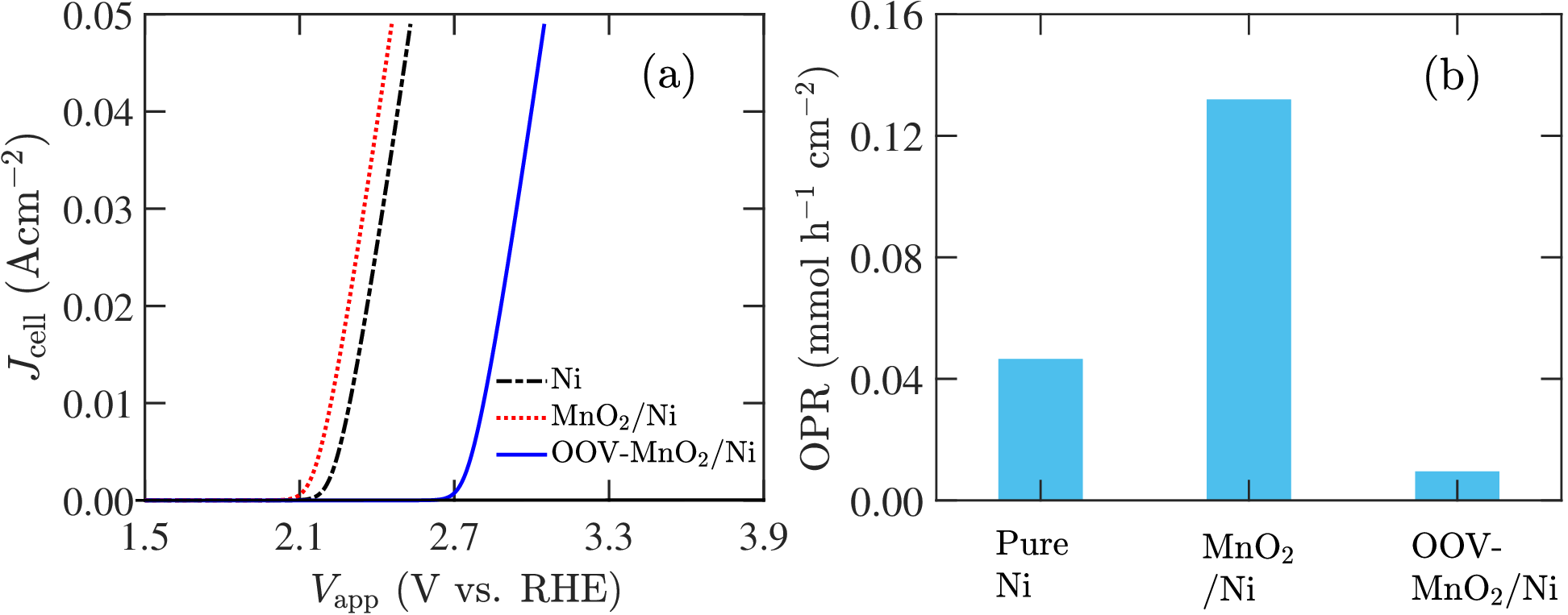}
    \caption{(a) Polarization curve (PC) and (b) oxygen production rate (OPR) for pure Ni(100) and the designed electrodes in oxygen evolution reaction (OER).}
    \label{Fig9}
\end{figure}
%
%
We also analyzed the PC and OPR of the designed electrodes during the OER process in electrochemical water splitting, as shown in Figs.~\ref{Fig9}(a) and (b), respectively. The MnO$_2$(110)/Ni(100) electrode exhibited superior PC characteristics compared to Ni(100) due to lower losses. However, the OOV-MnO$_2$(110)/Ni(100) demonstrated poor PC behavior because of significantly higher $\eta_{a,a}$ and $R_{\rm CT}$ during OER activity. The values of $V_{\rm on}$ were estimated to be 2.19 V for Ni(100), 2.12 V for MnO$_2$(110)/Ni(100), and 2.71 V for OOV-MnO$_2$(110)/Ni(100). The MnO$_2$(110)/Ni(100) achieved an OPR of 0.131 mmolh$^{-1}$cm$^{-2}$ at a $V_{\rm app}$ of 2.27 V, which was approximately three times greater than that of Ni(100). In contrast, we calculated the OPR for OOV-MnO$_2$(110)/Ni(100) at a $V_{\rm app}$ of 2.71 V, as 2.27 V was inadequate to initiate the OER process in electrochemical water splitting. Nevertheless, the calculated OPR for OOV-MnO$_2$(110)/Ni(100) remained significantly lower than that of Ni(100), indicating limitations that affect its effectiveness in efficient electrochemical systems.

The MnO$_2$(110)/Ni(100) structure serves as a bifunctional electrode, capable of functioning as both a cathode and anode in an electrochemical system. While it has slightly lower activity for HER, this electrode facilitates efficient water splitting to produce H$_2$ and O$_2$ with minimal energy losses. Additionally, it addresses instability issues that can occur at the electrode's surface. The designed electrodes with OVs demonstrated excellent electrocatalytic activity for HER and impressive overall electrochemical performance. However, they showed poor electrocatalytic activity for OER, which was attributed to the high concentration of OVs impacting the reconstruction of the OV sites in the presence of adsorbed O atoms. Notably, the OOV-MnO$_2$(110)/Ni(100) exhibited outstanding electrocatalytic activity and hydrogen production rates during HER process, comparable to those of noble metal-based electrodes. This suggests its potential use in electrochemical water splitting for efficient and effective hydrogen production.     

%

%
\section{Conclusion}
In summary, we designed a low-cost, high-performance Ni electrode using a unique catalyst made from a monolayer of $\beta$-MnO$_2$, focusing on OVs located at Bridge and Osub-top sites. Both the MnO$_2$(110)/Ni(100) structure with and without OVs exhibited excellent electronic DOS and charge properties, which ensured good metallic characteristics and enhanced electrical conductivity. For the MnO$_2$(110)/Ni(100) structure, we noted an increase in electrocatalytic activity for the OER, although there was a slight decrease in activity for the HER. In contrast, the MnO$_2$(110)/Ni(100) structure with OVs showed remarkable catalytic activity in HER. Specifically, the OOV-MnO$_2$(110)/Ni(100) exhibited an impressive $\Delta G_{\rm H}$ of $-0.015$ eV and $J_{0,c}$ of $10^{-0.774}$ Acm$^{-2}$, demonstrating superior performance compared to noble metal-based electrodes like Pt/C and Ir.

In our electrochemical analysis, OOV-MnO$_2$(110)/Ni(100) achieved an $\eta_{a,c}$ of $0.141$ V at $J = 10$ mAcm$^{-2}$ in 0.5 M H$_2$SO$_4$ solution, resulting in the HPR of 0.91 mmolh$^{-1}$cm$^{-2}$ at $ V_{\rm app} = 1.60$ V. However, the presence of OVs led to decreased electrocatalytic activity in the OER process due to their high concentration. Despite this slight reduction in HER activity, MnO$_2$(110)/Ni(100) can function as a bifunctional electrode, making it a promising option for use as both a cathode and anode in water-splitting applications, which facilitates efficient hydrogen production.   

%
%
\section*{Supplementary Material}
\noindent The supplementary material contains the calculated system energy, adsorption energy, and Gibbs free energy for hydrogen and oxygen at different sites of the electrodes.
\section*{Data Availability}
\noindent All data in the paper are present in the main text, which will also be available from the corresponding author upon reasonable request.
%
\section*{Author Declaration} The authors have no conflicts to disclose.

%

%
%
\small
\bibliographystyle{ieeetr}
\bibliography{references}

%
\end{document}